

\newbox\leftpage
\newdimen\fullhsize
\newdimen\hstitle
\newdimen\hsbody
\tolerance=1000\hfuzz=2pt
\overfullrule=0pt
\def\bigans{b }
\def\answ{b }
\ifx\answ\bigans\message{(this will come out unreduced.}


\magnification=1200\baselineskip=20pt
\font\titlefnt=cmr10 scaled\magstep3\global\let\absfnt=\tenrm
\font\titlemfnt=cmmi10 scaled\magstep3\global\let\absmfnt=\teni
\font\titlesfnt=cmsy10 scaled\magstep3\global\let\abssfnt=\tensy
\hsbody=\hsize \hstitle=\hsize 

\else\def\apans{h }
\message{(this will be reduced.}
\let\lr=l
\magnification=1000\baselineskip=12pt\voffset=-.31truein\hoffset=-.59truein
\hstitle=7truein\hsbody=3.4truein\vsize=9.5truein\fullhsize=7truein
\ifx\apansw\apans\special{ps: landscape}\hoffset=-.54truein
  \else\voffset=-.25truein\hoffset=-.45truein\fi
\font\titlefnt=cmr10 scaled\magstep4 \font\absfnt=cmr10 scaled\magstep1
\font\titlemfnt=cmmi10 scaled \magstep4\font\absmfnt=cmmi10 scaled\magstep1
\font\titlesfnt=cmsy10 scaled \magstep4\font\abssfnt=cmsy10 scaled\magstep1

\output={\ifnum\count0=1 
  \shipout\vbox{\hbox to \fullhsize{\hfill\pagebody\hfill}}\advancepageno
  \else
  \almostshipout{\leftline{\vbox{\pagebody\makefootline}}}\advancepageno
  \fi}
\def\almostshipout#1{\if l\lr \count1=1
      \global\setbox\leftpage=#1 \global\let\lr=r
   \else \count1=2
      \shipout\vbox{\hbox to\fullhsize{\box\leftpage\hfil#1}}
      \global\let\lr=l\fi}
\fi

%

\def\draftmode{\message{ DRAFTMODE }\def\draftdate{{\rm preliminary draft:
\number\month/\number\day/\number\yearltd\ \ \hourmin}}%
\headline={\hfil\draftdate}\writelabels\baselineskip=20pt plus 2pt minus 2pt
 {\count255=\time\divide\count255 by 60 \xdef\hourmin{\number\count255}
  \multiply\count255 by-60\advance\count255 by\time
  \xdef\hourmin{\hourmin:\ifnum\count255<10 0\fi\the\count255}}}

\def\title#1#2{\nopagenumbers\absfnt\hsize=\hstitle\rightline{}%
\centerline{\titlefnt\textfont0=\titlefnt%
\textfont1=\titlemfnt\textfont2=\titlesfnt #1}%
\centerline{\titlefnt\textfont0=\titlefnt%
\textfont1=\titlemfnt\textfont2=\titlesfnt #2
}%
\textfont0=\absfnt\textfont1=\absmfnt\textfont2=\abssfnt\vskip .5in}

\def\date#1{\vfill\leftline{#1}%
\tenrm\textfont0=\tenrm\textfont1=\teni\textfont2=\tensy%
\supereject\global\hsize=\hsbody%
\footline={\hss\tenrm\folio\hss}}
%

\def\nolabels{\def\eqnlabel##1{}\def\eqlabel##1{}\def\reflabel##1{}}
\def\writelabels{\def\eqnlabel##1{\hfill\rlap{\hskip.09in\string##1}}%
\def\eqlabel##1{\rlap{\hskip.09in\string##1}}%
\def\reflabel##1{\noexpand\llap{\string\string\string##1\hskip.31in}}}
\nolabels
%
\global\newcount\secno \global\secno=0
\global\newcount\meqno \global\meqno=1

\def\newsec#1{\global\advance\secno by1\xdef\secsym{\the\secno.}\global\meqno=1
\bigbreak\bigskip
\noindent{\bf\the\secno. #1}\par\nobreak\medskip\nobreak}
\xdef\secsym{}

\def\appendix#1#2{\global\meqno=1\xdef\secsym{#1.}\bigbreak\bigskip
\noindent{\bf Appendix #1. #2}\par\nobreak\medskip\nobreak}


\def\eqnn#1{\xdef #1{(\secsym\the\meqno)}%
\global\advance\meqno by1\eqnlabel#1}
\def\eqna#1{\xdef #1##1{(\secsym\the\meqno##1)}%
\global\advance\meqno by1\eqnlabel{#1$\{\}$}}
\def\eqn#1#2{\xdef #1{(\secsym\the\meqno)}\global\advance\meqno by1%
$$#2\eqno#1\eqlabel#1$$}

\global\newcount\ftno \global\ftno=1
\def\refsymbol{\ifcase\ftno
\or\dagger\or\ddagger\or\P\or\S\or\#\or @\or\ast\or\$\or\flat\or\natural
\or\sharp\or\forall
\or\oplus\or\ominus\or\otimes\or\oslash\or\amalg\or\diamond\or\triangle
\or a\or b \or c\or d\or e\or f\or g\or h\or i\or i\or j\or k\or l
\or m\or n\or p\or q\or s\or t\or u\or v\or w\or x \or y\or z\fi}
\def\foot#1{{\baselineskip=14pt\footnote{$^{\refsymbol}$}{#1}}\ %
\global\advance\ftno by1}


\global\newcount\refno \global\refno=1
\newwrite\rfile
\def\ref#1#2{${[\the\refno]}$\nref#1{#2}}
\def\nref#1#2{\xdef#1{${[\the\refno]}$}%
\ifnum\refno=1\immediate\openout\rfile=refs.tmp\fi%
\immediate\write\rfile{\noexpand\item{\the\refno.\ }\reflabel{#1}#2.}%
\global\advance\refno by1}
\def\addref#1{\immediate\write\rfile{\noexpand\item{}#1}}

\def\semi{;\hfil\noexpand\break}




\hyphenation{anom-aly anom-alies coun-ter-term coun-ter-terms}

 
\def\cF{\hbox{{$\cal F$}}}



\def\ba{{\bf a}}

\def\pmb#1{\setbox0=\hbox{#1}%
 \kern-.025em\copy0\kern-\wd0
 \kern .05em\copy0\kern-\wd0
 \kern-.025em\raise.0433em\box0 }



\def\cp #1 #2 #3 {{\sl Chem.\ Phys.} {\bf #1}, #2 (#3)}
\def\jetp #1 #2 #3 {{\sl Sov.\ Phys.\ JETP} {\bf #1}, #2 (#3)}
\def\jfm #1 #2 #3 {{\sl J. Fluid\ Mech.} {\bf #1}, #2 (#3)}
\def\jpa #1 #2 #3 {{\sl J. Phys.\ A} {\bf #1}, #2 (#3)}
\def\jcp #1 #2 #3 {{\sl J.\ Chem.\ Phys.} {\bf #1}, #2 (#3)}
\def\jpc #1 #2 #3 {{\sl J.\ Phys.\ Chem.} {\bf #1}, #2 (#3)}
\def\jsp #1 #2 #3 {{\sl J.\ Stat.\ Phys.} {\bf #1}, #2 (#3)}
\def\jdep #1 #2 #3 {{\sl J.\ de Physique I} {\bf #1}, #2 (#3)}
\def\macromol #1 #2 #3 {{\sl Macromolecules} {\bf #1}, #2 (#3)}
\def\pra #1 #2 #3 {{\sl Phys.\ Rev.\ A} {\bf #1}, #2 (#3)}
\def\prb #1 #2 #3 {{\sl Phys.\ Rev.\ B} {\bf #1}, #2 (#3)}
\def\pre #1 #2 #3 {{\sl Phys.\ Rev.\ E} {\bf #1}, #2 (#3)}
\def\prl #1 #2 #3 {{\sl Phys.\ Rev.\ Lett.} {\bf #1}, #2 (#3)}
\def\prsl #1 #2 #3 {{\sl Proc.\ Roy.\ Soc.\ London Ser. A} {\bf #1}, #2 (#3)}
\def\rmp #1 #2 #3 {{\sl Rev.\ Mod.\ Phys.} {\bf #1}, #2 (#3)}
\def\zpc #1 #2 #3 {{\sl Z. Phys.\ Chem.} {\bf #1}, #2 (#3)}
\def\zw #1 #2 #3 {{\sl Z. Wahrsch.\ verw.\ Gebiete} {\bf #1}, #2 (#3)}

\baselineskip=14pt
\hsize=14.8truecm
\vsize=23.5truecm

\def\cDR{{\cal DR}}
\def\cF{{\cal F}}

\def\cN{{\cal N}}

\def\cP{{\cal P}}
\def\cX{{\cal X}}
\def\cZ{{\cal Z}}

\def\om{\omega}

\def\halfspace{{\hskip0.4cm}}
\font\titlefont=cmbx10 scaled\magstep4

\centerline{\titlefont{Pattern formation in}}
\centerline{\titlefont{Laplacian growth:}}
\centerline{\titlefont{Theory}}

\font\titlefont=cmbx10 scaled\magstep1

\bigskip
\bigskip
\centerline{\titlefont{Raphael Blumenfeld}}

\centerline{Center for Nonlinear studies and Theoretical Division, MS B258}
\centerline{Los Alamos National Laboratory, Los Alamos, NM 87545, USA}

\bigskip
\bigskip
\bigskip
\item{}{\bf Abstract}
\smallskip
A first-principles statistical theory is constructed for the evolution of two
dimensional interfaces in Laplacian fields. The aim is to predict the pattern
that the growth evolves into, whether it becomes fractal and if so the
characteristics of the fractal pattern. Using a time dependent map the growing
region is conformally mapped onto the unit disk and the problem is converted to
the dynamics of a many-body system. The evolution is argued to be Hamiltonian,
and the Hamiltonian is shown to be the conjugate function of the real potential
field. Without surface effects the problem is ill-posed, but the Hamiltonian
structure of the dynamics allows introduction of surface effects as a repulsive
potential between the particles and the interface. This further leads to a
field representation of the problem, where the field's vacuum harbours the
zeros and the poles of the conformal map as particles and antiparticles. These
can be excited from the vacuum either by fluctuations or by surface effects.
Creation and annihilation of particles is shown to be consistent with the
formalism and lead to tip-splitting and side-branching. The Hamiltonian further
allows to make use of statistical mechanical tools to analyse the statistics of
the many-body system. I outline the way to convert the distribution of the
particles into the morphology of the interface. In particular, I relate the
particles statistics to both the distributions of the curvature and the growth
probability along the physical interface. If the pattern turns fractal the
latter distribution gives rise to a multifractal spectrum, which can be
explicitly calculated for a given particles distribution. A `dilute boundary
layer approximation' is discussed, which allows explicit calculations and shows
emergence of an algebraically long tail in the curvature distribution which
points to the onset of fractality in the evoloving pattern.

\bigskip
\vfill
\line{\hfil CNLS Newsletter 112, LALP-95-012, April 1995}
\eject

\newsec{Introduction}
\smallskip
\hsize=15.5truecm
\vsize=23.5truecm

Notwithstanding the abundance of phenomenological knowledge the morphologies of
 interfaces that grow in Laplacian fields are poorly understood theoretically.
In many cases these interfaces exhibit a rich variety of convoluted patterns.
Known examples are diffusion-limited aggregation (DLA), solidification of
supercooled liquid, electrodeposition and growth of bacterial colonies, to
mention but a few. At present there is no sufficient theoretical understanding
to allow for reliable predictions of the asymptotic patterns that such growths
evolve into, starting from the basic equations of motion (EOM). Existing
analyses are either of an effective medium type\ref\eff{L. A. Turkevich and H.
Scher, Phys. Rev. Lett. {\bf 55} 1026, (1985); E. Brener, H. Levine and Y. Tu,
{\it ibid.} {\bf 66} 1978 (1991)} or employ renormalisation group techniques
assuming similarity solutions and limit distributions. Recently two of the
latter generic approache managed to yield quite accurate values for the scaling
of the growth size (or the mass) with the linear size of the growth\ref\piet{L.
Piteronero, A. Erzan and C. Evertsz, Phys. Rev. Lett. {\bf 61}, 861 (1988),
Physica {\bf A151}, 207 (1988)}\ref\tch{T. C. Halsey and M. Leibig, Phys. Rev.
{\bf A 46}, 7793 (1992); T. C. Halsey, Phys. Rev. Lett. {\bf 72}, 1228 (1994)}.

A different direction to treat this problem was suggested exactly half a
century ago\ref\galin{L. A. Galin, Dokl. Akad. Nauk USSR {\bf 47}, 246 (1945);
P. Ya. Polubarinova-Kochina, Dokl. Akad. Nauk USSR {\bf 47}, 254 (1945); Prikl.
Math. Mech. {\bf 9}, 79 (1945)}. It was proposed that two-dimensional free
interfaces (i.e., in the absence of surface tension) be analysed by conformally
mapping the growing region onto the unit disk and studying the evolution of the
map rather than that of the interface. This idea was followed by
Richardson\ref\rich{S. Richardson, J. Fluid Mech., {\bf 56}, 609 (1972)} who
discovered that such a map enjoys a set of independent constants of motion.
Shraiman and Bensimon\ref\sb{B. Shraiman and D. Bensimon, Phys. Rev. {\bf A
30}, 2840 (1984)} took this issue a step further showing explicitly how the
problem can be converted to a many-body system and pointing out that the free
interface evolution is mathematically ill-posed. They demonstrated that the
formalism breaks down after a finite time due to instabilities with respect to
growth of perturbations along the interace on ever shorter lengthscales.
Without the curbing effect of surface tension, irregularities develop into cusp
singularities along the physical interface. There have been theoretical efforts
to counteract this catastrophic sharpening by using small surface tension to
cut off the short lengthscales in a renormalisable manner\ref\leo{D. Bensimon,
L. P. Kadanoff, S. Liang, B. I. Shraiman and C. Tang, Rev. Mod. Phys. {\bf 58},
977 (1986); W-s Dai, L. P. Kadanoff and S. Zhou, Phys. Rev. {\bf A 43}, 6672
(1991)}. These, however, met with another difficulty: Such an ad-hoc inclusion
turns out to constitute a singular perturbation to the EOM of the system. This,
in turn, means that solutions based on this approach are strongly sensitive to
initial conditions in the sense that very similar initial conditions can result
in completely different morphologies\ref\tanveer{S. Tanveer, Philosophical
Transactions of The Royal Society (London), {\bf A 343}, 155 (1993) and
references therein}. Yet, numerous observations show that the final
morpholgical properties are rather robust to the details of the growth and to
changes in initial conditions. A different approach has been proposed recently
to prevent formation of cusps by arguing that tip-splitting reduces high
surface curvatures. Implementing this idea into the EOM of the equivalent
many-body system results in production of particles\ref\bbi{R. Blumenfeld and
R. C. Ball,  Phys. Rev. {\bf E}, to appear}, as will be elaborated on in this
paper.

A significant question in this context, that has not been addressed much, is
whether the system supports a Hamiltonian or a Lyapunov function\ref\ba{R.
Blumenfeld, Phys. Lett. {\bf A 186}, 317 (1994)}\ref\sm{S. P. Dawson and M. B.
Mineev, Physica {\bf D73}, 373 (1994)}. This point is essential to a
fundamental understanding of this highly nonequilibrium growth process and
therefore to the thrust of this paper, as will become clear below. Although, as
formulated, the problem is known to support a set of conserved
quantities\rich\bbi\ref\mm{M. B. Mineev, Physica {\bf D 43}, 288 (1990)}, it is
unclear whether these can assist in finding a Hamiltonian for the system.

In this paper I first introduce the EOM of the interface. I show that the
interface follows Hamiltonian dynamics and that the Hamiltonian is directly
related to the physical two dimensional potential. I then derive the EOM of the
equivalent many-body system, whose particles are the zeros and poles of the
map. Using the existence of a Hamiltonian formulation, surface effects can be
described as a field that either repels particles from the surface or gives
rise to particles production. Both approaches overcome the ill-posedness of the
problem and extend its range of validity to infinite time. The Hamiltonian also
allows to analyse the statistical mechanics of the many-body system and in
particular the spatial distribution of the particles inside the unit disk. I
show how to extract information on the morphology of the interface from the
distribution of poles and zeros. This includes a calculation of the moments of
the growth probability distribution along the interface and the distribution of
the curvature. I demonstrate the calculation for the case when the particles
form a dilute gas near the unit circle and show that the distribution of the
curvature along the interface develops an algebraic tail, implying
non-negligible probabilities of formation of particularly sharp protrusions.
This tail points to the onset of fractality and self-similarity, a feature that
is not asumed a-priori. The third moment of the growth probability along the
interface is calculated explicitly in terms of the distribution of the
particles. This moment gives the fractal dimension of the pattern.

\smallskip
\newsec{The basic problem and the interface's EOM}
\smallskip
\hsize=15.5truecm
\vsize=23.5truecm

The fundamental problem of Laplacian growth can be formulated as follows.
Consider a Jordan curve, $\gamma(s)$, embedded in two dimensions which
represents the physical interface. This curve is parametrised by $0\leq s
<2\pi$, and is fixed at a given value of the potential field (electrostatic
potential for electrodeposition or concentration for diffusion controlled
processes). A higher potential is assigned to a circular boundary whose radius,
$R$, is very large compared to the growth size. The potential field, $\Phi$,
outside the area that is enclosed by $\gamma$ satisfies Laplace's Eq.
\eqn\Ai{\nabla^2 \Phi = 0\ .}
The interface is assumed to grow at a rate that is proportional to the local
gradient of the field which is normal to the interface\ref\fni{The theory
presented here can be generalised straightforwardly to the dielectric breakdown
model where the growth rate is proportional to $|\nabla\Phi|^{\eta}$}
\eqn\Aii{v_n = - {\bf \nabla}\Phi\cdot\hat{\bf n}\ .}
This rate is assumed to be sufficiently slow so that at any time the Laplacian
field can be considered to be static.
Denoting by $\zeta=x+iy$ the physical (complex) plane, we now conformally map
at each instant of time, $t$, the curve onto the unit circle (UC) in a
mathematical $z$ plane via
$\zeta = F(z,t)$.
The time-dependent interface is recovered from the map by
$\gamma(s,t)=\lim_{z\to e^{is}} F(z,t)$. The field gradient along the curve is
$-\nabla\Phi(\zeta) = -[\partial\Phi(\zeta)/\partial\zeta]^* = -i/(zF')^*$,
where $^*$ stands for complex conjugate and the prime indicates derivative with
respect to $z$. Using the fact that $z = e^{is} = 1/z^*$ on the interface,
Shraiman and Bensimon\sb derived the EOM for $\gamma$:
\eqn\Av{\partial_t\gamma(s,t) = -i\partial_s\gamma(s,t)
\Bigl[|\partial_s\gamma(s,t)|^{-2} + ig(s)\Bigr]\ .}
The first term within the square brackets on the r.h.s. of $\Av$ represents the
normal growth rate as constituted by relation $\Aii$. The second term, however,
is added by hand (but it is uniquely determined) and represents a tangential
velocity, or 'sliding', of a point along the interface. Denoting the entire
square brackets on the r.h.s. of Eq. $\Av$ as the limit of an analytic function
$G(z,t)$ (whose explicit form is unique and explicitly determinable - see
below), the EOM for $F(z,t)$ becomes
\eqn\Avi{\dot F = z F' G\ .}
For reasons to become clear below let us also write down the logarithimc
derivative of this equation with respect to $z$:
\eqn\Avia{{d\over{d t}}\ln F' = {1\over{F'}}{d\over{d z}}\left(z F' G\right)\
.}
The map needs in general to satisfy several constraints\bbi: First, we want the
topology of the boundary far away to remain unchanged under the map, which
enforces $\lim_{z\to\infty} F \sim z$. Second, the map must have no branch
cuts. These two conditions are satisfied by the following general form
\eqn\Avii{F' = A(t)\prod_{n=1}^N {{z - Z_n(t)}\over{z - P_n(t)}}\ ,}
where the time dependence appears in the scaling factor, $A$, and the locations
of the zeros and poles of $F'$. The first constraint imposes `charge
neytrality', i.e., the number of poles should equal the number of zeros. The
second constraint imposes a `dipolar neutrality', namely,
$\sum_nZ_n=\sum_nP_n$. It is possible to show that this form, and with a proper
choice of the number $N$, enables to describe {\it any} initial simply
connected cuspless curve that is allowed by this process. Therefore, this form
is quite general and not merely a small class of maps\ref\rb{N. Robidoux and R.
Blumenfeld, unpublished}.
Integrating $\Avii$ we have
\eqn\Aviii{\zeta = F(z,t) = A(t)\left[1 + \sum_{n=1}^N R_n \ln(z - P_n)
\right]\ .}
The quantities $R_n$ are the residues of the product in $\Avii$ at $P_n$
(without the prefactor $A$), and
$$\eqalign{Q_n &= 2\prod_{m=1}^N {{(1/Z_n - P_m^*)(Z_n-P_m)}\over
{(1/Z_n - Z_m^*)(Z_n - Z_{m'})}} \halfspace m'\ne n \cr
G &= G_0 + \sum_{n=1}^N {{Q_n}\over{z - Z_n}}\cr
G_0 &= \sum_{m=1}^N {{Q_m}\over{2 Z_m}} + \prod_{m=1}^N {{P_m}\over{Z_m}}\ .
\cr}$$
By contour-integrating around the location of the poles and zeros in $\Avi$ and
$\Avia$ one obtains their EOM:
\eqn\Ax{\eqalign{- A^2(t)\dot Z_n &= Z_n\Bigl\{G_0 + \sum_{m'} {{Q_n +
Q_{m'}}\over{Z_n - Z_{m'}}} \Bigr\} + Q_n\Bigl\{1 - \sum_m{{Z_n}\over{Z_n -
P_m}}\Bigr\} \equiv f^{(Z)}_n(\{Z\};\{P\}) \cr
- A^2(t)\dot P_n &= P_n\Bigl\{G_0 + \sum_m {{Q_m}\over{P_n - Z_m}}\Bigr\}
\equiv f^{(P)}_n(\{Z\};\{P\})\ . \cr }}
The behaviour of the particles under these these equations was discussed in
detail by Blumenfeld and Ball\bbi and will not be repeated here. From $\Avi$,
and using the requirement that there are no branch cuts, one finds that the
quantities $A(t)R_n(t)$ are {\it constants of the motion} that are independent
of each other and which are determined only by initial conditions. These
constants are directly related to those found by Richardson\rich and
mineev\mm. The time evolution of $A(t)$ is straightforwardly found from $\Avi$
\eqn\Axi{\dot A(t) = A(t) G_0\ .}
This relation shows how this prefactor dependends on the distribution of the
particles inside the unit disk. This rescaling factor is related directly to
the fractal dimension of the pattern, if it becomes fractal, as will be
discussed in section 5.

\smallskip
\newsec{The Hamiltonian structure}
\smallskip
\hsize=15.5truecm
\vsize=23.5truecm

Whenever a set of dynamical equations appears on the scene the first pertinent
question is whether it supports a Lyapunov function or a Hamiltonian. If it
does this gives access to a powerful bag of tools. I argue that the present
probcess is indeed Hamiltonian and further relate it to the actual field
$\Phi$. The first hint that the process is Hamiltonian comes from the physical
EOM $\Av$, when only normal growth is considered:
\eqn\Bi{\partial_t\gamma(s,t) = -i\partial_s\gamma(s,t) \left|
\partial_s\gamma(s,t)\right|^{-2} = -i{{\delta s}\over{\delta\gamma^*}}\ .}
This expression and its complex conjugate are equivalent to Hamilton's
equations. This relation suggests that the (rescaled) length of the actual
growth, $s$, may play a role of a Lyapunov functional, while $\gamma$ plays the
role of a field, whose real and imaginary parts are the canonical varialbles.
Although Eq. $\Bi$ is not formally correct this appealing interpretation begs
the question whether the EOM for the map, $F$, also supports such a structure.
It is not difficult to show\ref\mb{M. Mineev-Weinstein, private communication}
that this is indeed so: Multiply the complex conjugate of $\Av$ by
$\partial_s\gamma$ and rewrite its imaginary part in terms of the map\galin
$${\rm Im}\left\{F_t^* F_s\right\} = 1\ .$$
Writing $F=\om+i\chi$, with $\om$ and $\chi$ real functions, this equation
asserts that the Jacobian of the transformation from the coordinates $t-s$ to
the coordinates $\om-\chi$ is unity,
$\partial\left(\om,\chi\right)/\partial\left(t,s\right) = 1$.
It is then straightforward to show that when $z\to e^{is}$
\eqn\Biii{\dot \om = {{\partial s}\over{\partial \chi}}\halfspace ; \halfspace
\dot \chi = -{{\partial s}\over{\partial \om}} \halfspace {\rm and} \halfspace
\dot F = -i {{\partial s}\over{\partial F^*}} \ .}
Since $s = {\rm Im}\{\ln z\}$ Eq. $\Biii$ generalises to
\eqn\Bv{\dot F = -i{{\partial}\over{\partial F^*}} {\rm Im}\{\ln z\}=
-i{{\partial}\over{\partial F^*}} {\rm Im}\{\Phi\}\ ,}
where $\Phi$ is {\it the complex potential} in the physical plane. Thus the
trajectories of the system follow the stream lines, which are the conjugate of
the equipotential lines in the physical plane.

Yet another way to obtain a similar structure is as follows:
Define the complex function $\Psi \equiv F' + iz$. By manipulating Eq.
$\Avia$\ref\bii{R. Blumenfeld, Phys. Rev. {\bf E 50}, 2952 (1994)} one obtains
\eqn\Bvi{{{\partial \Psi}\over{\partial t}} = -i{{\partial H_0}\over{\partial
\Psi^*}}\ ,}
where $H_0 \equiv z F' G$, which appears on the r.h.s. of $\Avia$ is complex
and therefore does not enjoy a convenient translation into Hamilton's
equations.

The main point regarding the above arguments is not as much the exact form of
the Hamiltonian but rather that the p.d.e. that governs the interface's
evolution indeed follows Hamiltonian dynamics. In other words, given an initial
value of the Hamiltonian (the `energy') the system then follows a trajectory
that keeps this value constant. Since the many-body formulation is an
equivalent description of the growth process it follows that the system of
poles and zeros must also keep this quantity constant and hence the latter also
supports a Hamiltonian structure. To find the many-body Hamiltonian one inverts
relation $\Aviii$ to express $\ln z$ in terms of $F$ and then Im$\{\ln z\} =
\cF(F,F^*)$. I will not pursue this direction further here but rather argue
that the very existence of a Hamiltonian already paves the way to much
progress.

\smallskip
\newsec{Surface effects}
\smallskip
\hsize=15.5truecm
\vsize=23.5truecm

Turning to consider surface effects, it has already been mentioned that without
surface tension (or capilary forces) cusps form along the interface due to
instability of small corrugations\ref\ms{W. W. Mullins and R. F. Sekerka, J.
Appl. Phys. {\bf 34}, 323 (1963)}. But real growth processes clearly do not
admit cusps. In the cases that concern us here this is because the system has
to expend a macroscopic surface energy as the curvature increases. What does
the procedure of cusp formation correspond to in the many-body system? A local
protrusion (the incipient cusp) is caused, in the many-body system, by a zero
approaching the UC. Thus prohibition of high curvatures naturally corresponds
to keeping that zero from approaching the UC too closeely and hence to an {\it
effective repulsive potential} between the particles and the interface defined
by the UC. It should be stressed that only the existence of a Hamiltonian makes
it possible to use the term 'repulsive potential' with any proper meaning.

There are different ways to incorporate this idea into the theory: One is by
introducing a surface potential term in the many-body Hamiltonian\bii. Although
this may sound somewhat difficult since we don't know the exact Hamiltonian of
the many-body system, one can nevertheless insert such a term in the p.d.e.
$\Bvi$, $H = H_0 + V$, and derive the modified EOM for the particles. The
choice of the surface potential term determines the nature of the growth to a
large extent. The stronger the repulsion, the smoother the resulting
interfaces.  An example of a possible simple repulsive potential term is
\eqn\Ci{V = \sigma \lim_{z\to e^{is}} \ln \left[K(\{Z\},\{P\}) \right]\ ,}
where $K$ is the (complex) curvature in terms of the locations of the zeros and
the poles\bbi
\eqn\Cii{K(s,\{Z\},\{P\}) = \lim_{z\to e^{is}} |F'|^{-1} \Bigl\{ 1 +
\sum_{n=1}^N\bigl\{ {{Z_n}\over{z-Z_n}} - {{P_n}\over{z-P_n}} \bigr\} \Bigr\}\
,}
whose real part is the physical curvature. This particular surface potential is
simple in that it contributes a {\it constant} repelling term in the EOM of the
particles, $\Ax$. It has recently been shown that a term that diverges as a
particle approaches the UC would do better to describe the physics\bii.

Another approach is to make a deeper use of the fact that the particles move in
fact in a field. The field can have a vacuum that can accommodate particles and
antiparticles (zeros and poles). `Exciting' the vacuum (say, by fluctuations)
can then effect {\it creation and annihilation} of zeros and poles, a mechanism
that allows for tip-splitting and side-branching\bbi. But before considering
such a farfetching interpretation we need to convince ourselves that such a
picture is consistent with the present formalism and that it does not
contradict any of the basic premises. To this end consider again the derivative
of the map (Eq. $\Avii$) and rewrite it at the initial moment (say, $t_0$) in
the, seemingly redundant, form
\eqn\CCi{F'(z,t_0) = A(t_0) \prod_{n=1}^N {{z - Z_n(t_0)}\over{z - P_n(t_0)}}
\prod_{k=1}^\infty {{z - \Gamma_k(t_0)}\over{z - \Gamma_k(t_0)}}\ .}
The second product is unity at $t_0$ and at any time thereafter under the free
interface evolution because all its terms simply cancel out. It is
straightforward to see that, under the free-growth EOM, at any time $t>t_0$
$F'$ will retain this form  with only the position of the particles in the
first product and the value of $A$ changing. Consider now what happens when a
zero, $Z_i$, and a pole, $P_j$, collide. At the instant of collision the
particles occupy the same location and therefore their corresponding terms,
$(z-Z_i)/(z-P_j)$, cancel out in the first product on the r.h.s. of $\CCi$.
This is exactly an annihilation event. Such an event conserves the balance
between the numbers of zeros and poles, so that charge neutrality is not
violated. It also does not violate the dipolar neutrality, as can be
immediately verified. Turning to production events, the second product on the
r.h.s. of $\CCi$ can now be interpreted as a `vacuum' of pairs of zeros and
poles that do not manifest unless they are `excited'. One can envisage two
routes for this to occur: i) A fluctuation, of whatever origin, can virtually
separate such a pair; ii) A particle that moves with a high kinetick energy can
knock the pair apart. Particles with high velocities are usually zeros that are
close to the unit circle and therefore give rise to locally high curvatures.

So the form $\Avii$ is also naturally suited for creation. Immediately after a
zero-pole pair has been excited their locations are very close and it is quite
straightforward to verify from the EOM that such a close pair interact
repulsively, pushing apart and hence maintaining their identities. Differently
expressed, once created the particles are stable. Thus a pair, say at
$\Gamma_j$, can be excited into two individual particles at
$Z_j=\Gamma_j+\delta_1$ and $P_j=\Gamma_j+\delta_2$. Under production, as under
annihilation, the basic constraints need to be satisfied. Namely, an excitation
is in pairs for charge neutrality, and dipolar neutrality is staisfied by
imposing a relation between the locations of the excited particles (I should
mention that excitation by field fluctuations has to occur in quartets rather
than in pairs due to the dipolar constraint).
In a particular implementation of this idea Blumenfeld and Ball\bbi proposed
that excitation of a new pair is triggered by the proximity of a zero, say
$Z_n$, to the unit circle (scenario ii above). Such proximity leads to a high
curvature in front of $Z_n$ at $s={\rm arg}\{Z_n\}$ and the new zero-pole pair
is excited once the local curvature reaches a threshold value. To satisfy
dipolar neutrality the new pole occupies the location of $Z_n$ prior to the
production event, while the other two zeros are equidistantly and oppositely
situated around the pole. It turns out that the orientation of the zeros is not
limited by the constraints on the system and needs to be imposed following
another criterion. Blumenfeld and Ball introduced an energetic criterion along
the following lines: Recognising that surface energy increases with increasing
curvature, the location of the particles needs to minimise the local curvatute
at $s$. As it happens, this minimum corresponds to placing the two zeros in the
azimuthal direction about the location of the newly born pole. In the physical
plane, such an event constitutes exactly {\it tip-splitting}. Placing the zeros
along a radial ray equidistantly from the pole {\it maximises} the local energy
and corresponds to {\it side-branching} in the physical plane. Since the system
would rather minimise its local energy, these results suggest that
tip-splitting may have an {\it energetic}, rather than only stochastic, origin.
This particular procedure is not universal and other systems may follow other
criteria for particles production. It should be emphasised that it is quite
plausible that fluctuations of the field superimose on this mechanism and
stochastically induce tip-splitting and side-branching. Further studies in this
direction are currently being carried out and will not be elaborated on here.

\smallskip
\newsec{Noise and statistical analysis}
\smallskip
\hsize=15.5truecm
\vsize=23.5truecm

The foregoing assumed mostly a deterministic evolution. It is known\ref\how{S.
D. Howison, J. Fluid Mech. {\bf 167}, 439 (1986)} that the dynamical EOM of the
p.d.e. leads to a chaotic growth in the sense that if one starts from very
close initial conditions, one ends up very quickly with different deterministic
structure. This implies an efficient spread of the solutions in phase space
which, combined with the good fortune of having a Hamiltonian dynamics, ensures
the existence of a Gibbs measure. Namely, one expects to be able to construct a
partition function
\eqn\Di{\cZ = \int e^{-\beta H} \cDR\ ,}
where $\cDR\equiv\prod_{n=1}^N d^2 Z_n d^2 P_n$ is a an infinitesimal volume in
phase space and the Hamiltonian is that of the many-body system. The quantity
$\beta$ reflects the `noise' in the system and is obtained as a Lagrange
multiplier by imposing an average `energy' constraint on the distribution.
Expectation values of quantities such as the moments of the curvature and
moments of the field gradient, $|\nabla\Phi|$, can now be found via
\eqn\Dii{\langle X \rangle \equiv {\rm E}\{X\} = {1\over\cZ} \int X\ e^{-\beta
H} \cDR\ .}
Since, at present, the explicit form of $H$ is unknown, this approach is not
easy to implement.

Alternatively, we can construct a master equation for the evolution of the
distribution of the particles, $\cN(\{Z\},\{P\})$, using Liouville's theorem
\eqn\Diii{{{\partial\cN}\over{\partial t}} + \sum_{n=1}^N f^{(Z)}_n{{\partial
\cN}\over{\partial Z_n}} + f^{(P)}_n{{\partial \cN}\over{\partial P_n}} =
\Gamma \ ,}
where $\Gamma$ represents collisions and noise. To use Liouville's theorem let
me confine myself here to systems with conserved number of singularities (no
particles production). An extension to a nonconserved number of particles is
not difficult once a self-consistent renormalisation is introduced and will be
discussed elsewhere. Expecting a steady state distribution after rescaling the
growth by $A(t)$ we can discard the explicit time derivative. A solution of
this equation yields everything there is to know about the interface's
statistics. Unfortunately, as common in statistical mechanics, a general exact
solution is impossible. Rather than trying to solve this equation in some
approximation, let me demonstrate how such a solution can be converted into
information about the physical interface.
\smallskip
First, the distribution of the curvature, $\cP_K$, can be derived from $\cN$
using the relation
\eqn\Div{\cP_K = \int \cN\left(\{Z\},\{P\}\right) \delta\left\{K_1 -
{1\over{|F'|}}\left[1 + {\rm Re}\sum_{n=1}^N\left({1\over{1 - Z_ne^{-is}}} -
{1\over{1 - P_ne^{-is}}} \right) \right] \right\} \cDR \ ,}
where $K_1$ is the measurable curvature along the interface and $\delta$
denotes Dirac's delta-function. More generally, the distribution of any
morphology-related quantity, $\cX$, $\cP(\cX)$, that is expressible in terms of
the locations of the particles can be found in this manner. As an explicit
example of using the statistics, I now turn to calculate the values of the
moments of the growth probability distribution along the interface. This
calculation has been recently carried out by Blumenfeld and Ball\bbi and is
only briefly reviewed here. These moments play a central role in pattern
formation and growth, mostly because they were shown to lead to an
asymptotically stable multifractal function that is independent of initial
conditions or details of the growth. The probability that growth occurs at a
point $s$ along the interface at some time $t$ is proportional to the local
gradient of the field
$$p(s) = C_0 |\nabla\Phi(s,t)| = \lim_{z\to e^{is}} C_0 \left| F'(z,t)\right|
\halfspace ; \halfspace C_0 = 1/\oint_{\gamma(s,t)} |\nabla\Phi(l)| dl\ .$$
Therefore the quenched moments of this distribution are
\eqn\Wi{M_q = \oint_{\gamma} \left|\nabla\Phi\right|^q dl =
{1\over{2 \pi i}} \oint_{|z|=1} \left| F'(z)\right|^{1-q} {{dz}\over z} \ .}
Substituting from Eq. $\Avii$ we have
\eqn\Wii{M_q = {{A(t)^{1-q}}\over{2\pi i}} \oint \prod_{n=1}^N \left({{z -
Z_n}\over{z - P_n}}\right)^{{{1-q}\over 2}} \prod_{n=1}^N \left({{1 -z
Z_n^*}\over{1 - z P_n^*}}\right)^{{{1-q}\over 2}} {{d z}\over z}\ .}
The second product in the integrand, $\equiv J(z)^{(1-q)/2}$, is analytic
within the unit disk while the first contains $N$ poles of order $(1-q)/2$. For
$q>1$ these poles are located at the zeros of the map, while for $q<1$ the
poles of the integrand coincide with the poles of the map. In both regimes a
simple pole also exists at the origin. Thus it is straightforward to evaluate
this integral for odd values of $q$:
\eqn\Wiii{M_{q>1} = {1\over{A(t)^{2\nu}}} \left\{ \left(\prod_{n=1}^N
{{P_n}\over{Z_n}}\right)^{\nu} + {1\over{\nu !}} \sum_{n=1}^N
{{d^{\nu-1}}\over{dz^{\nu-1}}}\left[{{\left(z - P_n\right)^{\nu}}\over{z
J(z)^{\nu}}} \prod_{k \neq n} \left({{z - P_k}\over{z - Z_k}}\right)^{\nu}
\right]_{z=Z_n} \right\}}
where $\nu\equiv(q-1)/2$ is a positive integer number. Other values of $q>1$
can be obtained either by taking into account explicitly the branch-cuts that
are involved in the calculations, or by interpolating between the integer
moments. However, it is known that the knowledge of all the odd moments of the
probability density of a measure on a finite support is sufficient to determine
that probability density uniquely\ref\moments{B. Fourcade and A. -M. S.
Tremblay, Phys. Rev. {\bf B 36}, 8925 (1987); B. Fourcade and A. -M. S.
Tremblay, Phys. Rev. {\bf A 36}, 2352 (1987); A. Aharony, R. Blumenfeld, P.
Breton, B. Fourcade, A. B. Harris, Y. Meir and A. -M. S. Tremblay, Phys. Rev.
{\bf B 40}, 7318 (1989)}, and therefore the present calculation suffices. Of
particular interest is the third moment
\eqn\Wiv{M_3 = A(t)^{-2} \left[ \prod_{n=1}^N {{P_n}\over{Z_n}} + \sum_{n=1}^N
{{Q_n}\over{2 Z_n}} \right] = G_0/A^2(t) \ ,}
where the last step makes use of the definition of $G_0$ in section 2. Using
Eq. $\Axi$ we then obtain the exact relation between $M_3$ and $A(t)$,
$$M_3 = \dot A(t)/A^3(t)\ .$$
The third moment is directly related to the fractal dimension\ref\tch{T. C.
Halsey, Phys. Rev. Lett. {\bf 59}, 2067 (1987)}\ref\BB{R. C. Ball and R.
Blumenfeld, Phys. Rev. {\bf A 44}, R828 (1991)} by
$$D_f = \ln{M_3}/\ln{R_g}\ ,$$
where $R_g$ is the linear size (the radius of gyration or the radius of the
equivalent circular capacitor) of the growth. $D_f$ can be evaluated once the
distribution of the particles is known. Moreover, if indeed the pattern turns
fractal then this quantity should asymptope to a pure number. This is another
manifestation of the first-principles nature of the present approach: It is the
result that tells us whether the morphology becomes fractal without having to
assume such a solution from the outset.

Calculating for odd values of $q<1$ leads to an expression similar to $\Wiii$.
Since all these quantities depend on the locations of the particles one can
find their distribution over many growth realisations via an integral similar
to $\Div$. An interesting observation is the following: Comparing the
expressions for the positive and negative moments shows that these enjoy
exactly the same form with the location of the zeros interchanged with the
locations of the poles. This, combined with measurements on DLA that show a
{\it distinct difference} between the negative and positive moments of the
growth probability distribution implies qualitatively different spatial
distribution of the two species of particles. A partial confirmation of this
conclusion can be indeed observed in numerical calculations where the
trajectories of the poles and the zeros display markedly different
behaviours\bbi.

 \smallskip
\newsec{The dilute boundary layer approximation}
\smallskip
\hsize=15.5truecm
\vsize=23.5truecm

To gain insight into some features of the morphology let me now discuss briefly
the distribution of the curvatures using a dilute boundary layer approximation
for the distribution of the particles.
The motivation behind this approximation is as follows: From Eq. $\Cii$ one can
observe that the curvature is dominated by the zeros that are closest to the
UC. Namely, considering a zero, $Z_n=(1-\rho)e^{is_n}$ with $\rho\ll 1$, the
curvature in front of this zero is (see Eq. $\Cii$)
\eqn\Dv{K_1(s_n) \approx C/\rho^2\halfspace ,}
where, to a good approximation, $C$ is independent of either $\rho$ or the
locations of the other particles. A similar consideration for a pole near the
UC shows that the local curvature is {\it regular} in $\rho$. We can therefore
neglect the contribution of poles in the following approximation. Since these
are the zeros with $\rho \ll 1$ that dominate the curvature we consider only
the zeros within a ring $1-\rho_c < |z| <1$ and assume that their density is
dilute and isotropic (the isotropy assumption can in fact pertain only to a
discrete number of global arms). It can be easily shown that this corresponds
to analysing the exposed parts of the physical growth. By assumption then the
curvature at $s_n$ depends mainly on the radial location of $Z_n$, whose
distribution is $\cP_n(\rho)$. Thus
\eqn\Dvi{\cP_K = \cP_n(\rho) d\rho/dK = Const.\ K_1^{-3/2}
\cP_n\left((C/K_1)^{1/2}\right)\ .}
In the absence of a solution to the master equation $\Diii$ we have no
information on $\cP_n$. Nevertheless, we can observe that even if this
distribution is well behaved (i.e., all its relevant moments are finite) the
distribution of $K_1$ exhibits {\it algebraically decreasing tails} for high
curvatures. This immediately suggests the onset of fractality, which
generically originate from such tails. It is important to note that nowhere
along the construction of the theory did we assume fractality or
self-similarity. Yet, an algebraic tail appears which can easily generate such
a structure. The moments of $K$ are dominated by the long tail and satisfy
$\mu_q \sim K_{max}^{q-1/2} \sim a^{1/2-q}$, where $a$ is a small cutoff
lengthscale. With growth the cutoff radius reduces by a factor of $1/A(t)$ and
therefore
$\mu_q \sim A(t)^{q-1/2}$.
It is quite plausible that $\cP_n$ also introduces algebraic tails, directly
affecting the behaviour of $\cP_k$. A more accurate analysis is only possible
once we have a solution to Eq. $\Diii$ and attempts in this direction are being
carried out.

Before concluding this section I should mention that it is possible, using this
approximation, to also evaluate the aforementioned moments of the growth
probability, $M_q$\bbi. Such an evaluation shows that these moments depend
strongly on the {\it negative} moments of the distribution of zeros' distances
from the UC. If indeed the patterns becomes multifractal, such an evaluation
should yield the left hand side of the multifractal spectrum (the regime
governed by the exposed parts of the growth) directly.

\smallskip
\newsec{Discussion and concluding remarks}
\smallskip
\hsize=15.5truecm
\vsize=23.5truecm

To conclude, a first-principles theory for two dimensional Laplacian growth was
described. The growth was shown to be Hamiltonian with trajectories of constant
energy corresponding to the stream lines in the two-dimensional physical field.
This was argued to indicate that the equivalent many-body system is also
Hamiltonian. Although not discussed here, an extension of this formalism to a
continuous density of zeros and poles is possible and has been carried out.\bii
The resulting nonlocal dynamical equations, however, currently seem too complex
for an analytical treatment. Surface energy was shown to lead to a field that
acts on the particles in the mathematical plane. The field concept could be
incorporated either as a repulsive potential between the particles and the unit
circle or by giving rise to particles production. The latter corresponds to
tip-splitting and side-branching. Thus this seems a natural mechanism to
prevent cusp formation along the physical interface and connect surface effects
to tip-splitting. I discussed the distribution of the zeros and poles inside
the unit disk and related it explicitly to the morphology of the interface. The
main advantage of this approach is that the nonequilibrium growth is
describable in statistical mechanics formalism, the language of equilibrium
phenomena. The distribution of the curvature was analysed and an exact
calculation of the moments of the growth probability distribution along the
interface was discussed. To address the fundamental issue of onset of
scale-invariance and self-similarity, I used an approximation which showed that
the curvature distribution develops an {\it algebraic tail}, which naturally
gives rise to a fractal pattern. It is this author's belief that this approach
and its possible generalisations are very promising as a framework for
constructing theories for the morphology of growth processes in other
dimensionalities and in fields that satisfy equations other than Laplace's.

\vfill\immediate\closeout\rfile
\baselineskip=18pt\centerline{{\bf REFERENCES}}\bigskip\frenchspacing%
\input refs.tmp\vfill\eject\nonfrenchspacing
\bye